\documentclass[aps,prb,twocolumn,showpacs]{revtex4-1}
\usepackage{graphicx}
\usepackage{amsmath,amsfonts}
\usepackage{color}

\def\beq{\begin{equation}}
\def\eeq{\end{equation}}
\def\ba{\begin{align}}
\def\ea{\end{align}}
\newcommand{\er}{\mathbf{r}}
\newcommand{\ee}{{\rm e}}
\newcommand{\nR}{n_{\rm R}}

\newcommand{\gc}{g_{\rm C}}

\begin{document}

\title{Phase ordering kinetics of a nonequilibrium exciton-polariton condensate}

\author{Micha\l{} Kulczykowski and Micha\l{} Matuszewski}
\affiliation{Instytut Fizyki Polskiej Akademii Nauk, Aleja Lotnik\'ow 32/46, 02-668 Warsaw, Poland}

\begin{abstract}

We investigate the process of coarsening via annihilation of vortex-antivortex pairs, following the quench to 
the condensate phase in a nonresonantly pumped polariton system. 
We find that the late-time dynamics 
is an example of universal phase ordering kinetics, 
characterized by scaling of correlation functions in time. 
Depending on the parameters of the system, the evolution of the characteristic length scale $L(t)$
can be the same as for the two-dimensional XY model,
described by a power law with the dynamical exponent $z\approx 2$ and a logarithmic correction, or
$z\approx 1$ which agrees with previous studies of conservative superfluids.

\end{abstract}


\maketitle

\section{Introduction}

One of the major achievements of statistical physics is
the ability to describe complex systems of many particles
with a limited set of variables describing their collective behavior.
The universality of phase transitions is a particularly striking example of such
reduction, where the multitude of physical models is divided
into a finite number of universality classes characterized by
certain symmetry properties and critical scaling laws. 
Phase transitions from disordered to ordered states are often accompanied
by the creation of defects, such as domain walls, vortices or strings~\cite{KibbleZurek}.
In most realistic situations, these defects  subsequently decay in time,
and the system  undergoes gradual phase ordering or coarsening~\cite{Bray_PhaseOrderingKinetics}.
In this nonequilibrium, late-time stage of dynamics, physical systems frequently
exhibit universality characterized by a single length scale $L(t)$
that dictates the temporal evolution of all relevant quantities, 
such as correlation functions. The knowledge of symmetries and 
the character of the dominant coarsening process
is sufficient to determine the evolution of this length scale. 
The theory of universal coarsening has been successfully applied to a wide variety of systems,
from metallurgy and phase separation of fluids~\cite{Cahn,Furukawa_Scaling}
to biological systems and opinion dynamics~\cite{Castellano_StatisticalSocial}. 

Recently, the classical concept of phase ordering kinetics was extended
to the quantum realm in studies of atomic Bose-Einstein condensates~\cite{Sachdev_PhaseOrdering,Moore_Spinor,Sarma_Coarsening,Svistunov}.
In both the spinless~\cite{Sachdev_PhaseOrdering} and spinor superfluid 
gases~\cite{Moore_Spinor,Sarma_Coarsening,Kawaguchi_DomainGrowth,Kawaguchi_FerromagneticCoarsening,Blakie_CoarseningSpin1}
links with the corresponding classical systems were established.
In the area of quantum fluids of light~\cite{Carusotto_QuantumFluids,Polaritons,Kasprzak_BEC}, 
spontaneous creation of vortices during nonadiabatic exciton-polariton condensation was observed in~\cite{Deveaud_VortexDynamics}
and investigated theoretically in the context of the Kibble-\.Zurek 
mechanism~\cite{Matuszewski_UniversalityPolaritons,Liew_InstabilityInduced}. 
Vortex dynamics was a topic of many studies, eg.~\cite{Vina_VorticesCoherently,Bramati_AngularVortexChain,
Deveaud_HydrodynamicVortices,Deveaud_VortexStreets,Fraser_VortexAntivortex,Rubo_WarpingVortices,Skolnick_InteractionsOnVortices,
Szymanska_TOPOVortices,Yamamoto_VortexPair,Whittaker_TriggeredVortices,Deveaud_SelectivePhotoexcitation,Baumberg_GeometricallyLocked},
and the process of vortex-antivortex annihilation was observed experimentally in~\cite{Vina_VortexAntivortex,Sanvitto_AllOpticalControl,Bramati_MergingVortices}.
However, to date the universal coarsening dynamics has not been investigated.

Here, we verify the scaling hypothesis in the model of a nonresonantly pumped polariton 
condensate~\cite{Wouters_ExcitationSpectrum,Bobrovska_Adiabatic,Szymanska_NonequilibriumBKT,Sanvitto_TopologicalOrder}. 
We consider two sets of parameters such as material constants, pumping power etc. We find examples of universal phase ordering
with complete collapse of correlation functions after rescaling spatial coordinates by the length scale $L(t)$.
The length scale evolves according to a power law with the exponent $z$ depending on the parameters.
In the first case, $z\approx 2$ with a logarithmic correction, as predicted previously 
for two-dimensional vector or complex fields with purely diffusive dynamics~\cite{Bray_PhaseOrderingKinetics,Pargellis_DefectDynamics}.
In the second case, the dynamical scaling of $L(t)$ is found
to be the same as determined previously for conservative superfluids~\cite{Sachdev_PhaseOrdering,Blakie_CoarseningSpin1}, 
with $z\approx 1$. This shows that polariton systems can display various types of universal dynamics, 
which can be achieved by modifying the material parameters of the sample.

It is important to note that in this work we consider the time-evolving properties of a system that has suddenly crossed a phase 
transition, and not the critical properties of the phase transition itself. 
The latter have been intensively investigated in both recent experimental~\cite{Yamamoto_PNAS,Yamamoto_AlgebraicBKT,Sanvitto_TopologicalOrder} 
theoretical~\cite{Sieberer_DynamicalCritical,Szymanska_NonequilibriumBKT} works. 
It has been claimed that polariton systems display a kind of dissipative Berezinskii-Kosterlitz-Thouless (BKT) transition, while 
critical exponents may differ from the ones obtained in thermal equilibrium~\cite{Yamamoto_PNAS,Yamamoto_AlgebraicBKT,Szymanska_NonequilibriumBKT,Sanvitto_TopologicalOrder}. Here, we assume that the system is sufficiently far away from the critical point on the ordered side of it, so the system converges to an approximately defect-free phase.

The idea of universal coarsening dynamics is grounded on the scaling hypothesis,
which states that at late times there is a single characteristic length scale describing the 
large-scale features of the system~\cite{Furukawa_Scaling,Bray_PhaseOrderingKinetics}.
The configuration of  defects remains unchanged in time, 
in the statistical sense, if the spatial coordinates are 
scaled by this length scale which usually grows according to a power law $L(t)\sim t^{1/z}$. Here $z$ is
the nonequilibrium dynamical exponent, which is in general different from the dynamical {\it critical} exponent
of the phase transition which may have produced the defects in the first place~\cite{KibbleZurek,Biroli_IsingAnnealing}.
Consider, for instance, the first order, equal-time correlation function.
It follows that the following scaling holds
\begin{align}
\label{g1}
g^{(1)}({\bf d},t) &= \frac{1}{N}\int\langle \psi^*(\er,t) \psi(\er+{\bf d},t) \rangle {\rm d}\er =\\ \nonumber
&= f(d/L(t)).
\end{align}
with $d=|{\bf d}|$ and $f(0)=1$, where $N=\langle \int{|\psi|^2} {\rm d}\er \rangle$. 
While the scaling hypothesis has been rigorously proven only in several cases, numerical studies indicate
its validity in many physical systems~\cite{Bray_PhaseOrderingKinetics}.

The value of the exponent $z$ depends in general on the dimensionality of the system, the character of the
coarsening processes (diffusive, inertial, etc.), symmetries, and conservation laws~\cite{Bray_PhaseOrderingKinetics}. 
For non-conserved scalar fields, 
such as the Ising model or model A of diffusion-reaction~\cite{Hohenberg_DynamicalCriticalPhenomena}, it takes the value $z=2$.
In the case of conserved scalar fields the coarsening is slower with $z=3$,
which can be understood as the effect of the reduced number of accessible intermediate states~\cite{Moore_Spinor}.
When the transport is inertial rather than diffusive,
faster scaling with $z=3/2$ is predicted~\cite{Furukawa_Droplet,Kawaguchi_DomainGrowth}.

In the case of vector or complex fields, the existence of topological defects 
often dominates the phase ordering dynamics. This leads to different values 
of $z$, and in some cases, to logarithmic corrections, with the notable case of 
the two-dimensional XY model displaying the $L(t)\sim (t/\ln t)^{1/2}$ 
dependence~\cite{Pargellis_DefectDynamics,Yurke_CoarseningXY,Bray_XYModel,BrayBlundell_OnModel,Jelic_QuenchXYModel}.
For the subclass of diffusive models, arguments based on the comparison of the local and 
global energy change allow for the prediction of the growth laws in the
general case~\cite{Bray_EnergyScalingApproach,Bray_GrowthLaws}.
Renormalization group methods are also useful, however they fail to predict the logarithmic 
corrections~\cite{Bray_PhaseOrderingKinetics,Bray_EnergyScalingApproach}, 
which may turn out to be significant~\cite{Bray_XYModel}.

\section{Model}

We consider the wave function of nonresonantly pumped polariton condensate $\psi({\bf x},t)$ coupled to the reservoir described
by a density field $\nR({\bf x},t)$
~\cite{Wouters_ExcitationSpectrum, Wouters_ClassicalFields}
\begin{align}
\label{GPE}
i\textrm{d}\psi=&\left[ -\frac{\hbar D}{2m^*}\nabla^2+\frac{g_C}{\hbar}|\psi|^2+\frac{g_R}{\hbar}n_R+\right.\\\nonumber
&\left.+\frac{i}{2}\left(Rn_R-\gamma_C\right)
\right]\psi\textrm{d}t+\textrm{d}W,\\ \nonumber
\frac{\partial n_R}{\partial t}=&P-\left(\gamma_R+R|\psi|^2\right)n_R,
\end{align}
where $P$ is the exciton creation rate determined by the external optical or electrical pumping, $m^*$ 
is the effective mass of lower polaritons, 
$\gamma_{\rm C}$ and $\gamma_{\rm R}$ are the polariton and reservoir loss rates, 
and $R$ is the rate of stimulated scattering from the reservoir to the condensate, and $g_{\rm C}$, $g_{\rm R}$ are
the rates of repulsive polariton-polariton and reservoir-polariton interactions, respectively.
We also introduced $D=1-iA$ with $A$ being a small constant accounting for the 
energy relaxation in the condensate~\cite{Bobrovska_Stability,Sieberer_DynamicalCritical,Bloch_ExtendedCondensates,Bloch_GapStates}.
Alternatively, one may introduce a complex coefficient in front of the time derivative term~\cite{Wouters_Superfluidity,Wouters_EnergyRelaxation}.
The above phenomenological model has been successful in describing a number of 
different experimental situations in exciton-polariton 
condensates~\cite{Deveaud_VortexDynamics,Yamamoto_VortexPair}.
The complex stochastic noise $\textrm{d}W$, corresponding to disturbance associated with particles incoming and leaving the condensate, 
can be obtained within the truncated Wigner approximation~\cite{Wouters_ClassicalFields}
\begin{align}
\label{noise-psi}
\langle\textrm{d}W({\bf x})\textrm{d}W^*({\bf x}^{\prime})\rangle&=\frac{\textrm{d}t}{2(\Delta  x)^d}(Rn_R+\gamma_C)\delta_{{\bf x},{\bf x}^{\prime}}
,\\ \nonumber
\langle\textrm{d}W({\bf x})\textrm{d}W({\bf x}^{\prime})\rangle&=0.
\end{align}
In the absence of noise, a spatially uniform solution is given by $\psi({\bf x},t)=\psi_0e^{-i\mu_0t}$ and $n_R({\bf x},t)=n_R^0$.
Above the threshold pumping $P>P_{\rm th}=\gamma_C\gamma_R/R$ a stable condensate exists with the condensate density $|\psi_0|^2=(P/\gamma_C)-(\gamma_R/R)$ and $\mu_0=g_C|\psi_0|^2+g_Rn_R^0$. %, $n_R^0=\gamma_C/R$, and $\mu_0=g_C|\psi_0|^2+g_Rn_R^0$.
We define the healing length, which corresponds to the size of the vortex core as $\xi=2 \pi \hbar / \sqrt{m \gc |\psi_0|^2}$.
The scaling hypothesis may be valid only when typical distance between vortices is much larger than their size,
i.e.~the condition $L(t)\gg \xi$ must be fulfilled.

\begin{figure}
\includegraphics[width=8.5cm]{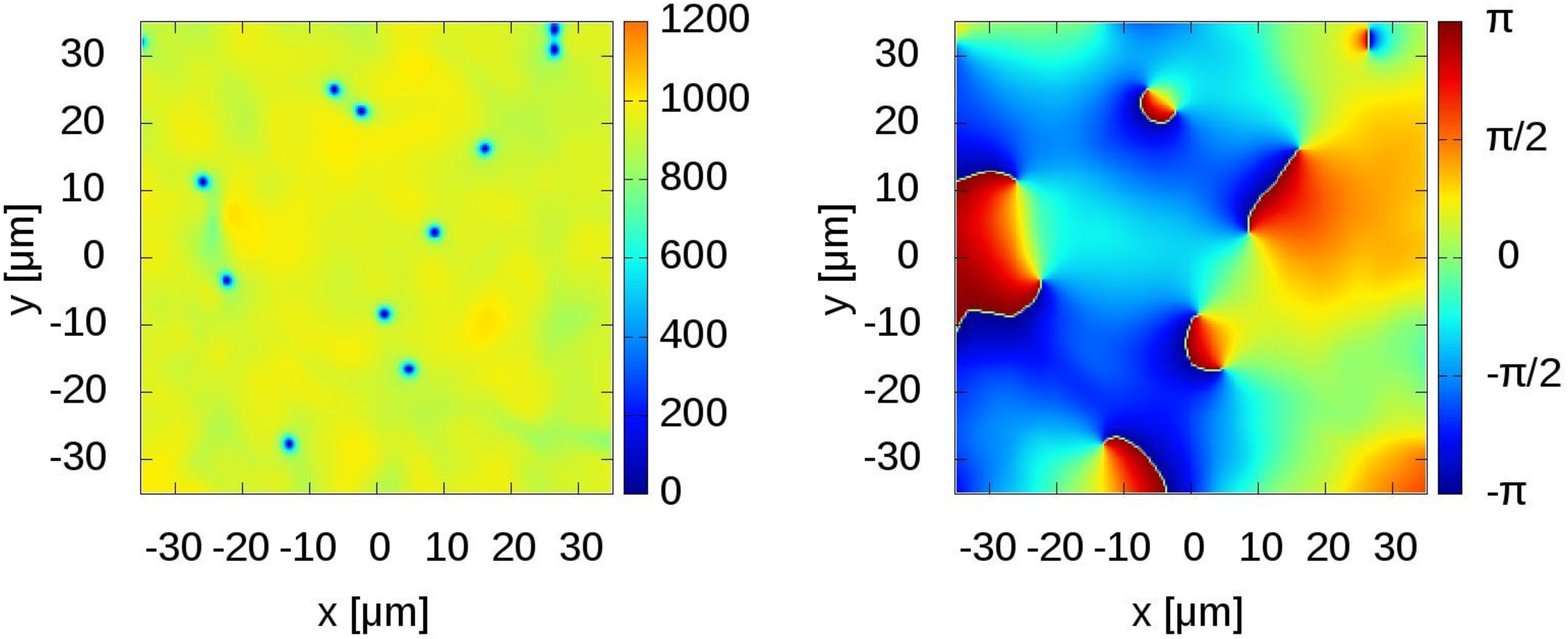}
\includegraphics[width=8.5cm]{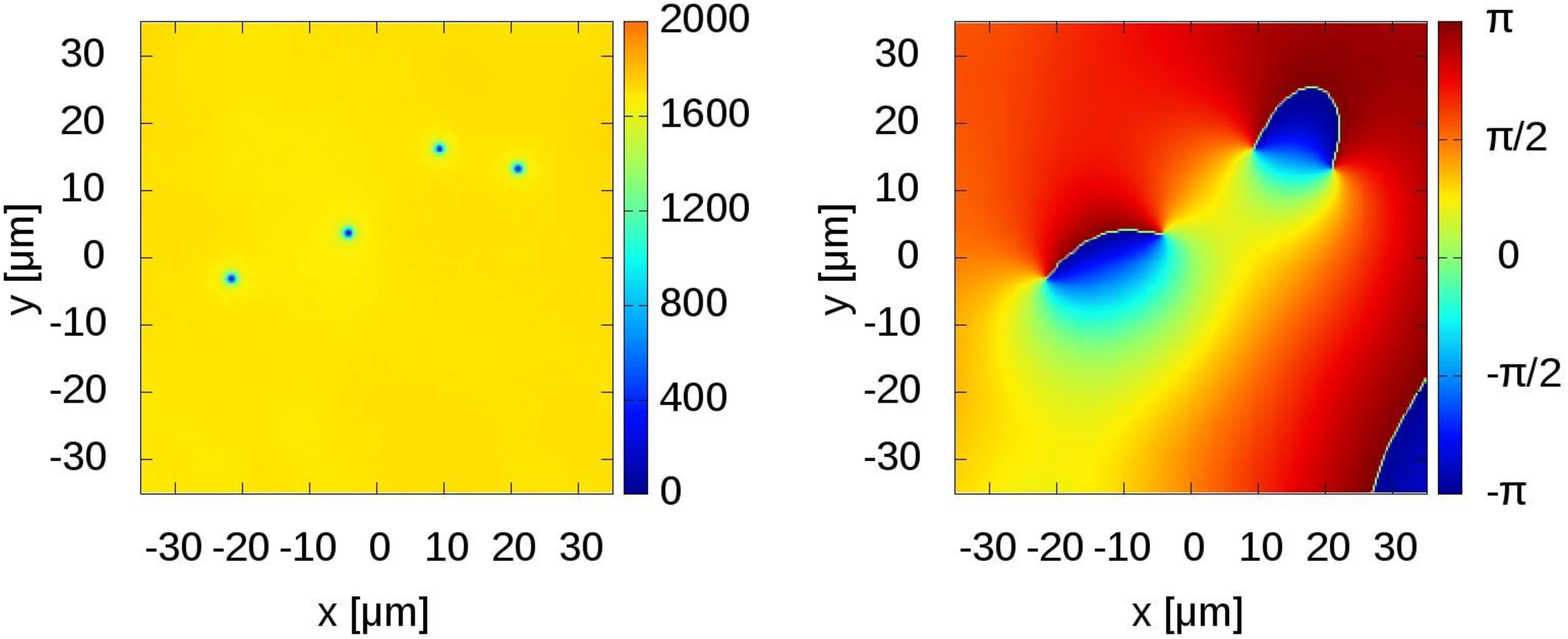}
\caption{Phase ordering through annihilation of vortex-antivortex pairs.
Density (left column) and phase (right column) of the condensate wave function at $t=50\,$ps (top) and $t=150\,$ps (bottom) after the quench. 
Parameters are  $m^*=5\times 10^{-5} m_{\rm e}$, 
$\gamma_C^{-1}=50\,$ps, $\gamma_R^{-1}=8\,$ps, $A=0$, $g_C=3.4\,\mu$eV$\mu$m$^2$, $g_R=7.2\,\mu$eV$\mu$m$^2$, 
$R=5.5\times 10^{-3}\,\mu$m$^2$ ps$^{-1}$, $P=40\,\mu$m$^{-2}$ ps$^{-1}$, $\xi=2.9 \mu$m.  }
\label{fig:snapshots}
\end{figure}

\begin{figure}
\includegraphics[width=6.5cm]{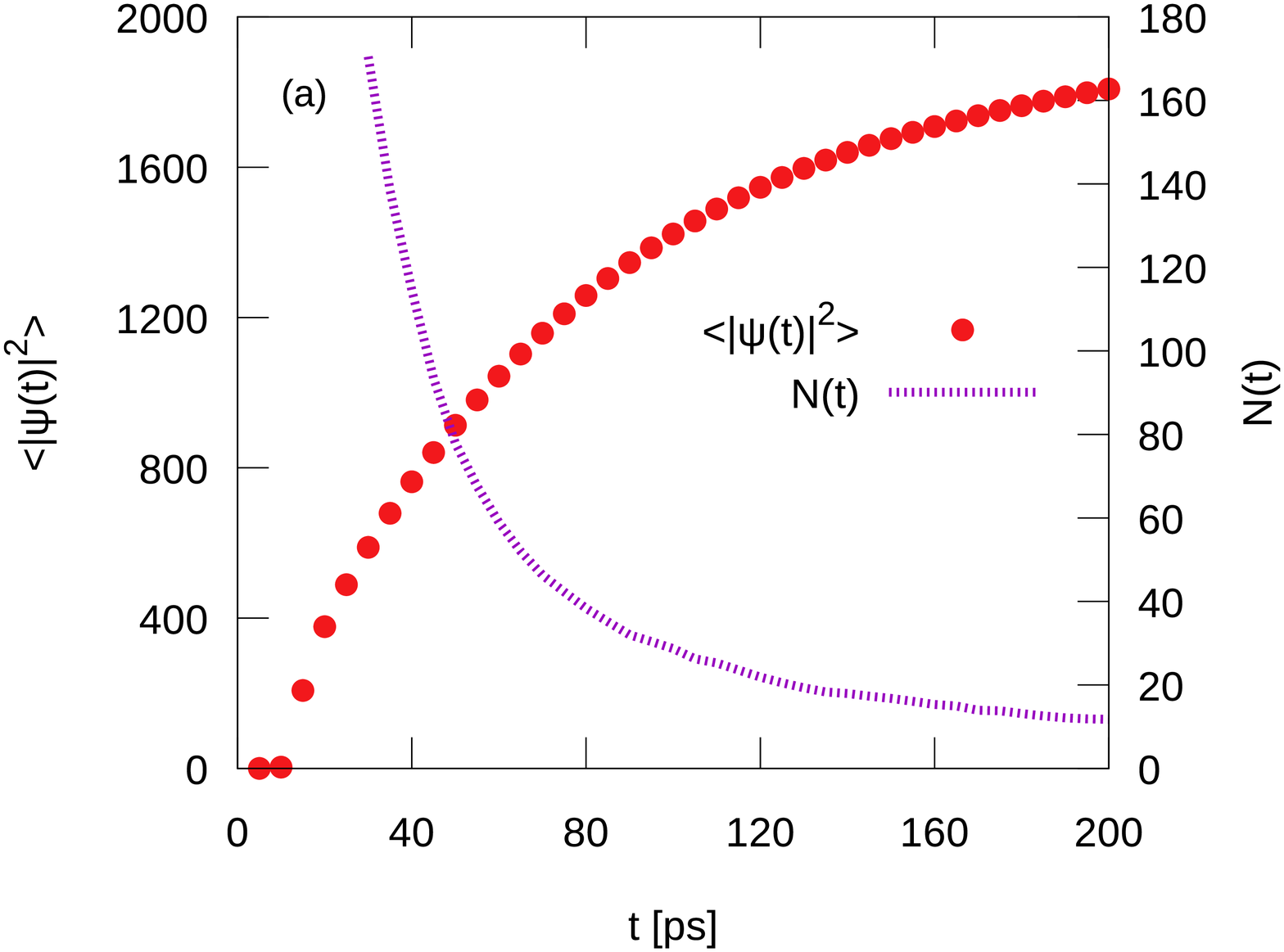}
\includegraphics[width=6.5cm]{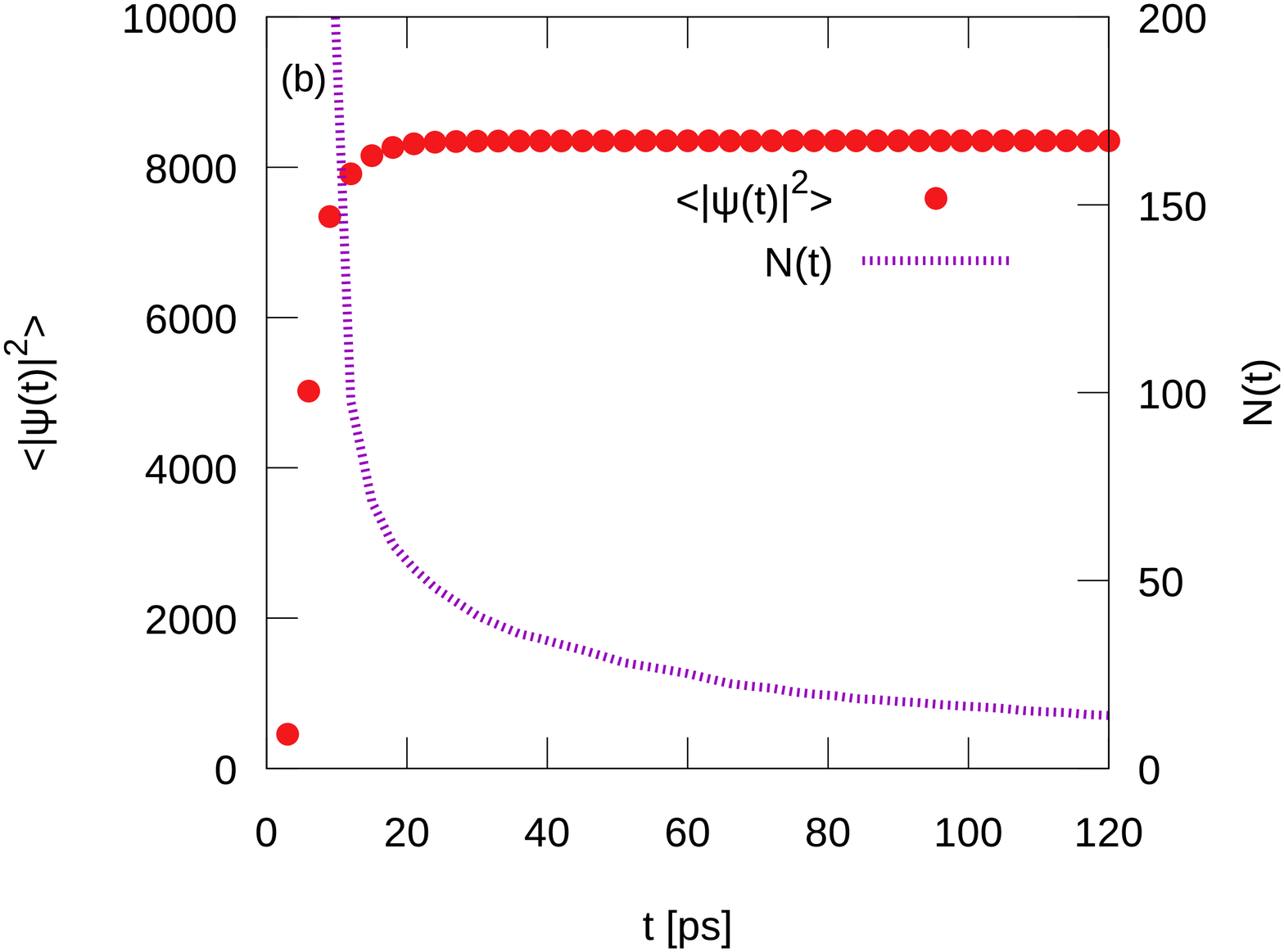}
\caption{Typical evolution of the mean condensate density and number of vortices.
  (a) The pair annihilation occurs in parallel with the saturation of the density.
(b) The evolution after $t\gtrsim 20\,$ps corresponds to pure phase ordering kinetics. In 
this case the density of defects decays according to the scaling law~(\ref{XYlaw}). Parameters in (a) are the same
as in Fig.~\ref{fig:snapshots}, while in (b) we use $\gamma_C^{-1}=3.3\,$ps, $\gamma_R^{-1}=3\,$ps, $A=0.1$,
$g_C=3\,\mu$eV$\mu$m$^2$, $g_R=6\,\mu$eV$\mu$m$^2$, 
$R=2.3\times 10^{-4}\,\mu$m$^2$ ps$^{-1}$, $P=3\times 10^3\,\mu$m$^{-2}$ ps$^{-1}$, $\xi=1.6 \mu$m. Averaged over 16 realizations of the Wigner noise.
}
\label{fig:evolution}
\end{figure}

\section{Results}

To investigate the process of phase ordering, we solved Eqs.~(\ref{GPE}) numerically 
on a rectangular mesh with size $l=150\,\mu$m with periodic boundary conditions. Starting form an empty
initial condition, $\psi, \nR=0$, 
the emergence of a polariton condensate from Wigner noise for $P>P_{\rm th}$ is accompanied by the spontaneous creation
of phase vortices in the process analogous to the Kibble-\.Zurek 
mechanism. The detailed description of this process was presented in~\cite{Matuszewski_UniversalityPolaritons,Liew_InstabilityInduced}.
Here, however, we are not interested in the process of defect creation, but rather in the
long-time dynamics of coarsening, which occurs when the defects are already established. 
We find that, analogously as in the case of the two-dimensional 
XY model~\cite{Pargellis_DefectDynamics,Yurke_CoarseningXY,Bray_XYModel,BrayBlundell_OnModel,Jelic_QuenchXYModel,Kawaguchi_FerromagneticCoarsening},
it occurs predominantly via annihilation of vortex-antivortex pairs. 

To illustrate this, in Fig.~\ref{fig:snapshots} 
we present snapshots of the amplitude and phase of the condensate wave function. The number of vortex-antivortex pairs
decreases monotonously.
In Figure~\ref{fig:evolution}, we display the evolution of the
condensate density and the number of vortices in function of time. The number of vortices is estimated from the number of 
points on the numerical mesh where the wave function is approximately zero and phase winding occurs, at a specific time.
Early in the evolution this number is very large as we start from a disordered state at low density, but
actual vortices become well established only when the density becomes large.

The results are displayed for two sets of parameters, 
corresponding to two situations that can occur in the system. In Fig.~\ref{fig:evolution}(a), 
the pair annihilation is effective already at the stage of the dynamics when the condensate density is not yet
fully established due to slow saturation of the density. 
This means that the defect creation and phase ordering overlap temporally, and a 
clear distinction between the two processes is not possible, similar as 
in~\cite{Zurek_WhenSymmetryBreaks2D,Biroli_IsingAnnealing,Jelic_QuenchXYModel,Deutschlander_KZMColloids}. 
On the other hand, if parameters of the system are chosen such that they correspond to a lower quality sample, 
with a shorter polariton lifetime, the stationary density is established 
more quickly, and the dynamics for $t\gtrsim 20\,$ps 
is practically purely due to phase ordering, as shown in Fig.~\ref{fig:evolution}(b). The parameters used throughout the paper are summarized in Table~\ref{table}.

\begin{figure}
\includegraphics[width=6.5cm]{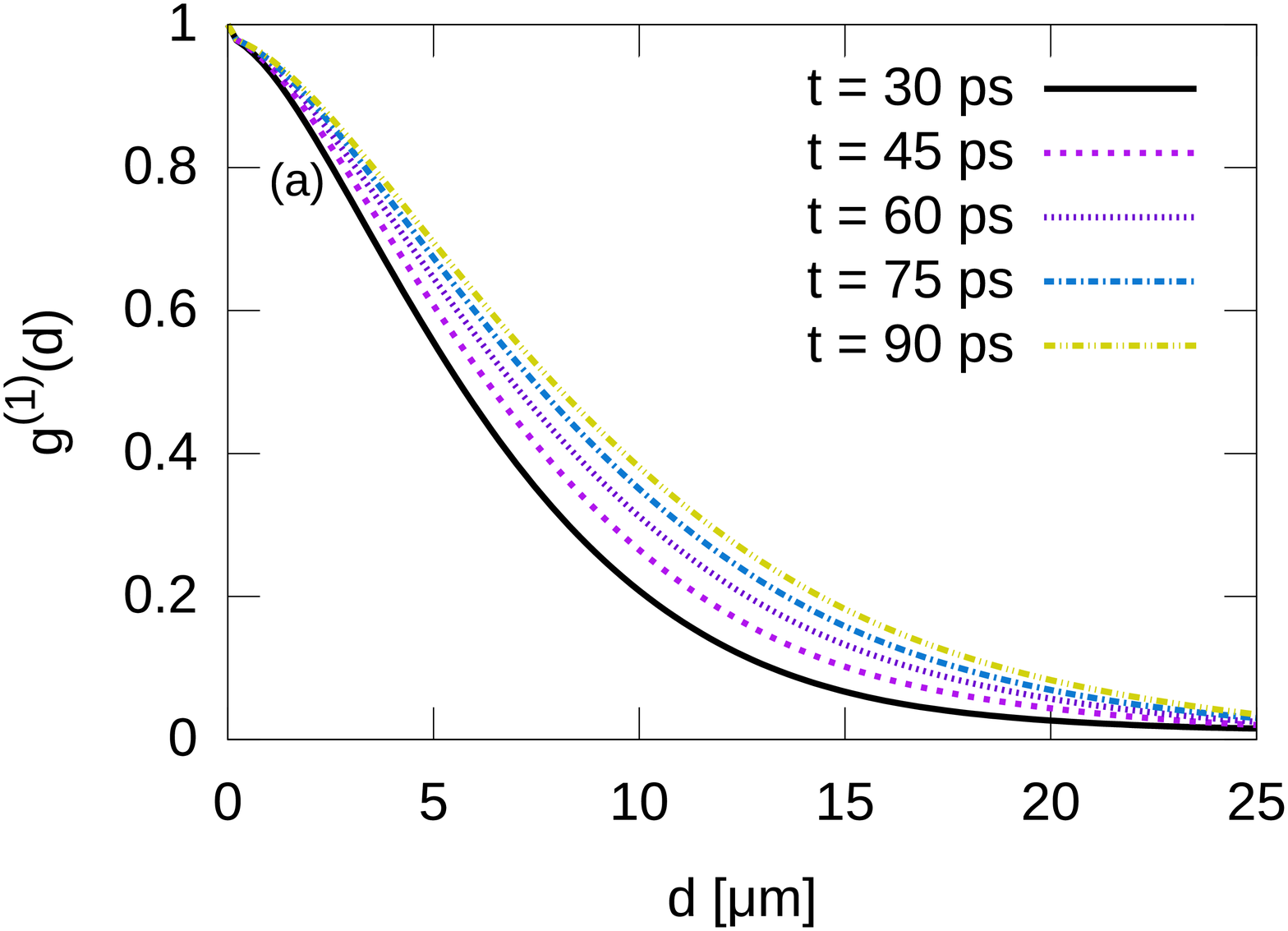}
\includegraphics[width=6.5cm]{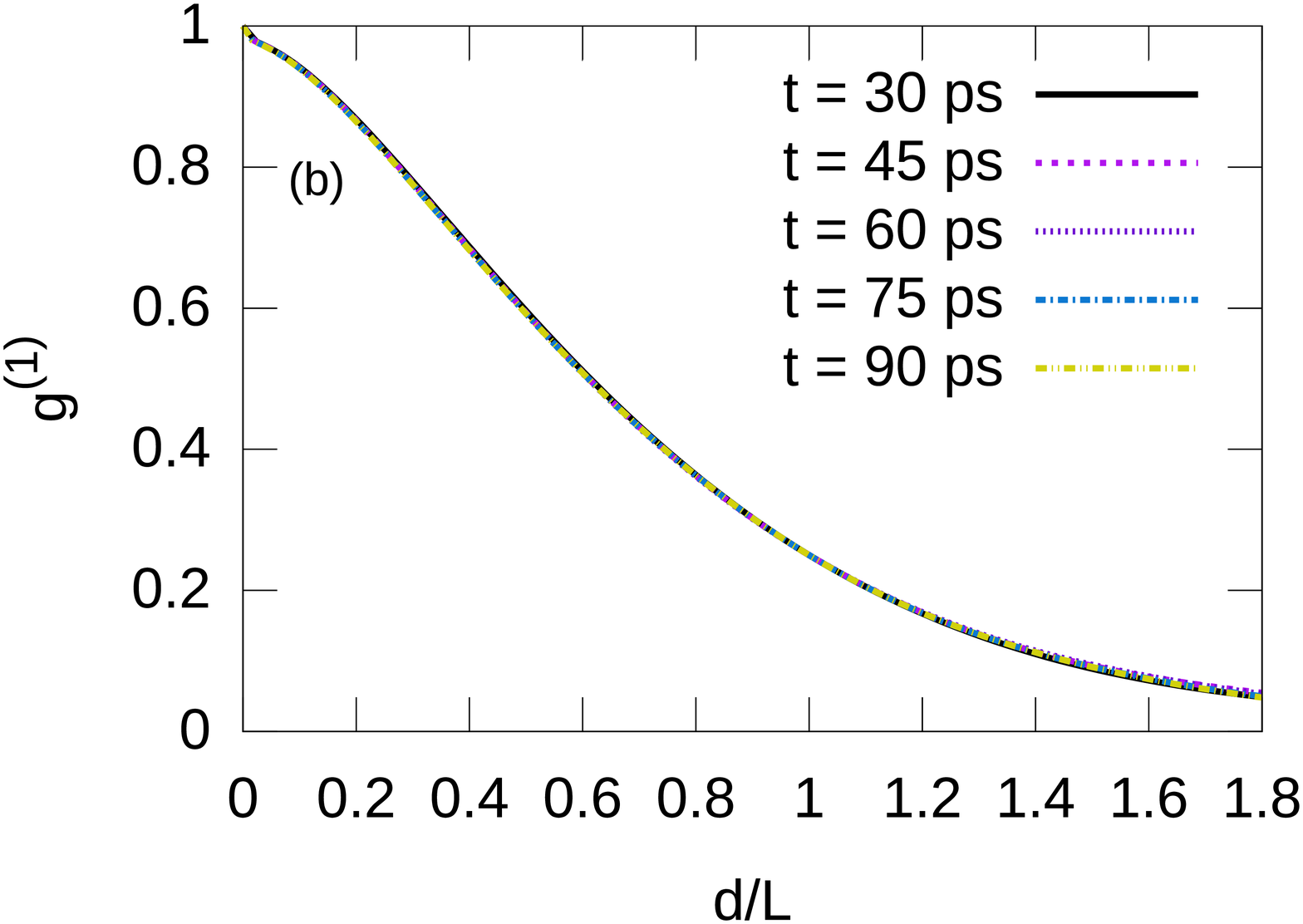}
\caption{(a) First-order correlation function vs.~distance, at several evolution times. The increase
in spread of $g^{(1)}$ is a result of  coarsening, and an increase of the characteristic length scale $L(t)$,
which we define as the value of distance $d$ at which $g^{(1)}=0.25$. (b) Collapse of the correlation function
after rescaling the $d$ axis by $L(t)$, confirming the scaling hypothesis. The parameters are the same as in Fig.~\ref{fig:evolution}(b).
}
\label{fig:g1}
\end{figure}

A more complete information about the statistical properties of the system is given by the correlation functions.
We confirm the scaling hypothesis by directly verifying the scaling property of the first order correlation function~(\ref{g1}).
In Figure~\ref{fig:g1}(a) we show $g^{(1)}(d)$ plotted at several instants of time during the pure phase ordering 
stage, averaged over 16 realizations
of the truncated Wigner simulations. As an estimate the length scale
$L(t)$ in~(\ref{g1}) we choose the value of $d$ for which the condition $g^{(1)}(d) = 0.25$ is fulfilled. We note 
that in contrast to the case of scalar fields, where the correlation function often exhibits an oscillatory tail,
in the present case there are no oscillations, which is generally the case if sharp domains walls are 
absent~\cite{Bray_XYModel,BrayBlundell_OnModel,Blakie_CoarseningSpin1}. 
We obtain a perfect collapse for the scaled correlation function $f(d/L(t))$, 
which confirms that the scaling hypothesis is valid in this case, see Fig.~\ref{g1}(b). 

\begin{figure}
\includegraphics[width=6.5cm]{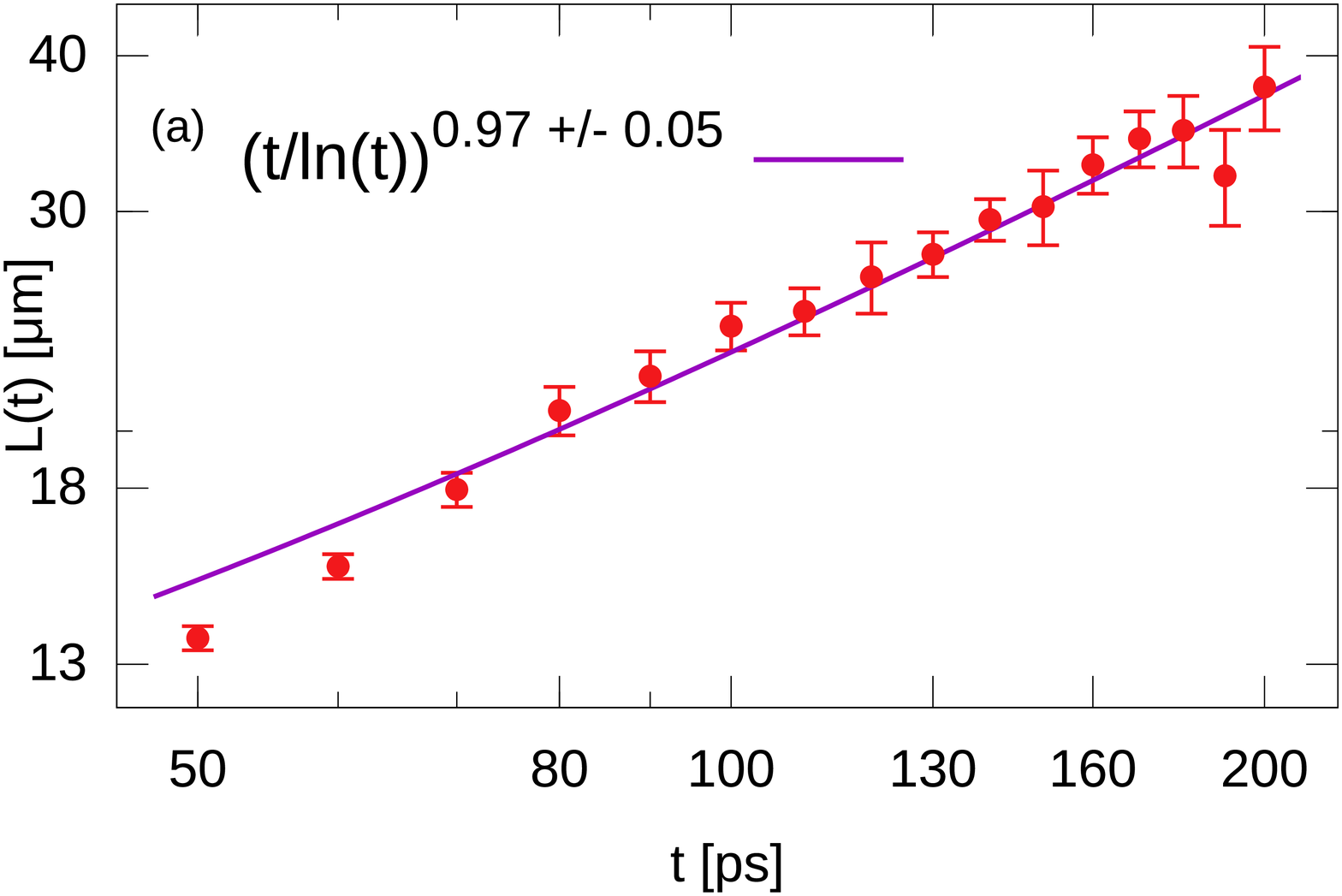}
\includegraphics[width=6.5cm]{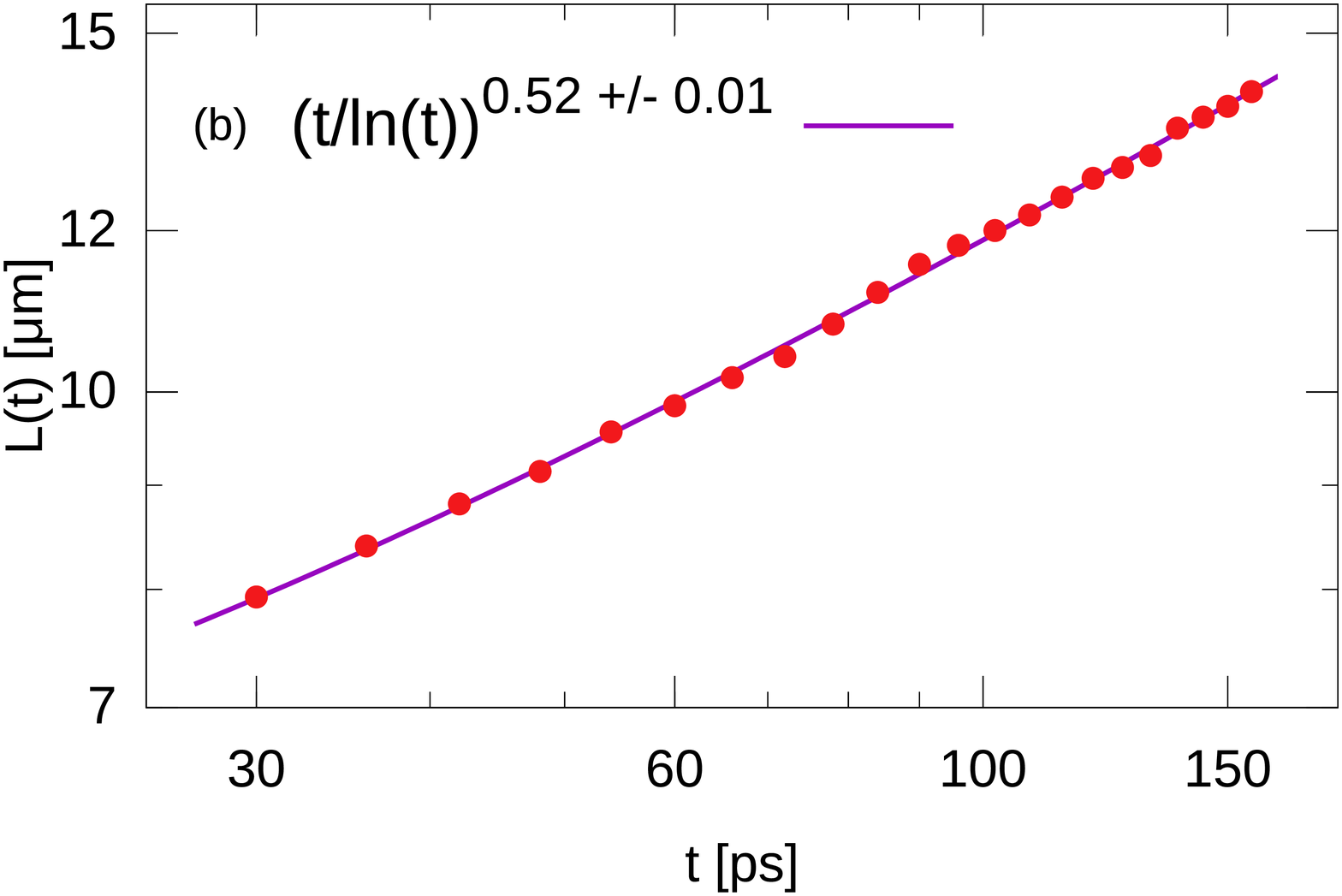}
\caption{(a) Time dependence of the length scale $L(t)$ in
the case of Fig.~\ref{fig:evolution}(a). The dynamical exponent
attains the value $z\approx 1$, in agreement with previous studies of conservative 
superfluids~\cite{Sachdev_PhaseOrdering,Blakie_CoarseningSpin1}. 
(b) The case of Fig.~\ref{fig:evolution}(b) with pure phase ordering. In this case the length scale follows the
universal scaling law for vector systems in two dimensions with nonconserved order parameter, Eq.~(\ref{XYlaw}) with $z\approx 2$. Error bars correspond to the standard deviation of the estimation of $L(t)$, determined from the values of $L(t)$ that vary from one realization to another. The error bars in (b) are comparable to the size of a single point. 
The number of averaged realizations was 30 in (a) and 150 in (b).
}
\label{fig:scalings}
\end{figure}

The verification of the scaling hypothesis allows one to expect a particular form of the scaling law for the 
time dependent length scale $L(t)$. We find that in the ``clean'' case of Fig.~\ref{fig:evolution}(b) 
it follows closely the scaling law predicted for two-dimensional systems with a vector order parameter
\beq \label{XYlaw}
L(t) \sim \left(\frac{t}{\ln (t/t_0)}\right)^{1/z},
\eeq
with $z\approx 2$, as shown in Fig.~\ref{fig:scalings}(b). 
In particular, it is the same as in the case of the
XY model in two dimensions~\cite{Pargellis_DefectDynamics,Yurke_CoarseningXY,Bray_XYModel,BrayBlundell_OnModel,Jelic_QuenchXYModel}.
We note that the logarithmic correction stems from the existence of a second relevant length scale and is absent
if the initial conditions contain no free vortices~\cite{Bray_XYModel}.
This scaling law is different from the one predicted for conservative atomic condensates 
both in the spinless~\cite{Sachdev_PhaseOrdering} and spinor cases~\cite{Blakie_CoarseningSpin1}, 
where $z\approx 1$ was obtained. 
Nevertheless, an atomic condensate model including the effects of dissipation~\cite{Kawaguchi_FerromagneticCoarsening} predicted $z=2$. 
This highlights the crucial difference between the conservative and dissipative systems from the point of view of coarsening 
and demonstrates that in the case of polariton condensates dissipation is essential. 
%% It is interesting to note that
%% this system belongs to the same universality class as the nonconservative 
%% XY model~\cite{Jelic_QuenchXYModel} but differ crucially from conservative atomic
%% condensates which are described by formally very similar Gross-Pitaevskii equations.

The above scaling law with $z=2$ can be explained by the balance between the vortex-antivortex attractive force  
and the effective friction~\cite{Bray_EnergyScalingApproach,Kawaguchi_FerromagneticCoarsening,Bray_XYModel}. 
Consider an isolated (anti)vortex of the form $\psi=A(r,t)\ee^{\pm i\phi(r,t)-i\mu_0 t}$. 
When the dynamics is diffusive~\cite{Wouters_ExcitationSpectrum,Szymanska_NonequilibriumCondensation}, 
far from the vortex core $A\approx|\psi_0|$ and the evolution of phase is given by the Kardar-Parisi-Zhang 
equation~\cite{Diehl_TwoDimensionalSuperfluidity,Bobrovska_Adiabatic}.
In the $L(t)\rightarrow +\infty$ limit 
this equation reduces to $\partial \phi / \partial t \approx -(1/\Gamma) \delta H / \delta \phi$, 
where $H$ is the nonlinear sigma model Hamiltonian corresponding to the kinetic part 
of~(\ref{GPE}).
The energy of the vortex is divergent as $E_v\sim\ln (l/a)$, where $l$ is the system size 
and $a\approx \xi$ is the ``microscopic'' cutoff~\cite{Fetter_vortices,Bray_XYModel}. 
For a vortex-antivortex configuration, $l$ is replaced by $R$, the distance between the vortices.
From the pair energy we obtain the attractive force $F=-dE/dR\sim 1/R$. The energy dissipation for 
a vortex moving with velocity $v$  can be calculated as
$d E / dt = \int d^2r (\delta H / \delta \phi)  (\partial \phi / \partial t) \sim
-\int d^2r (\partial \phi / \partial t)^2 = -v^2 \int d^2r (\partial \phi / \partial x)^2 \sim -v^2 E_v$, with the friction
constant $\gamma\sim E_v\sim \ln (R/a)$. The evolution of the average distance between pairs is $d R / dt \sim F/\gamma \sim 1/R \ln (R/a)$, which results in $R(t) \sim (t/\ln (t/t_0))^{1/2}$. 

We also investigate in more detail the case of the second set of parameters from Fig.~\ref{fig:evolution}(a).
As we mentioned before, here the phase ordering dynamics is not pure, but takes place when 
the stationary state density is yet to be established. We find that although 
in this case the scaling hypothesis does not precisely describe the system dynamics, 
some quantitative predictions can still be formulated about the phase ordering process. 
Indeed, we find only slight deviations from the collapse of the $g^{(1)}$ correlation functions (not shown),
which is related to the existence of a second time scale corresponding to the slow saturation of density and not to the coarsening.
In Fig.~\ref{fig:scalings}(a) we show the time evolution of the length scale in this case.
The data fits well to the theoretical prediction~(\ref{XYlaw}) with $z\approx 1$, in agreement
with the results on vortex driven coarsening in ferromagnetic atomic condensates 
in the easy-axis configuration~\cite{Blakie_CoarseningSpin1}.
This result is further supported by the value of exponent $z=1$ for model E of superfluid helium~\cite{Hohenberg_DynamicalCriticalPhenomena},
and the estimated value $z\approx 1.1$ for conservative spinless condensate~\cite{Sachdev_PhaseOrdering}. We note
that in this regime the ratio of the average distance between the vortices to the vortex size is
larger, $L(t)/\xi=17.2$ (at $t=150$ ps) as compared to the previous case where $L(t)/\xi=4.5$ (at $t=90$ ps). The possible transition between the two scalings in function of this parameter would require more detailed numerical study with longer evolution times and computational box sizes.

We note that at large distances, the force between vortex and antivortex may become repulsive~\cite{Diehl_Duality}, which could lead to slowing down of the annihilation, and the saturation of the length scale L(t) at late times. However, we did not observe such behaviour for the parameters that were considered.

\begin{table}[hbt]
\renewcommand{\arraystretch}{1.2}
\begin{tabular}{| c | c | c | }    
\hline
Figures & 1(a,b), 2(a) and 4(a) & 2(b), 3(a,b) and 4(b)\\
\hline
A & 0 & 0.1 \\
\hline
$g_C$ & $3.4 \mu eV\mu m^2$ & $3 \mu eV\mu m^2$ \\
\hline
$g_R$ & $7.2 \mu eV\mu m^2$ & $6 \mu eV\mu m^2$ \\
\hline
$\gamma^{-1}_{C}$ & $50 ps$  & $3.3 ps$  \\
\hline
$\gamma^{-1}_{R}$ & $8 ps$ & $3 ps$ \\
\hline
R & $5.5 \times 10^{-3} \mu m^2$  & $2.3 \times 10^{-4} \mu m^2$ \\
\hline
P & $ 40 \mu m^{-2} ps^{-1}  $ & $ 3 \times 10^3 \mu m^{-2} ps^{-1}  $ \\
\hline
$\xi$ & $2.9 \mu m$  & $1.6 \mu m$ \\
\hline
\end{tabular}
\caption{Parameters used in simulations to obtain the data presented in the Figures.}
\label{table}
\end{table}

In conclusion, we confirmed that universal phase ordering can occur in exciton-polariton condensates.
We found a scaling regime corresponding to purely diffusive dynamics and the one which is similar as in systems 
with conservative dynamics, while in the latter case the precise physical interpretation is not clear. We note that this is not the only system to display various universal behavior in different parameter regimes; 
in binary liquids diffusive, viscous hydrodynamic and 
inertial hydrodynamic regimes exist with different values of critical exponents~\cite{Furukawa_Droplet,Bray_PhaseOrderingKinetics}. 
To our best knowledge polariton condensates are unique in that the transition
is between the scaling laws determined by the dynamics of topological defects.

%% Finally, we note that while our study is limited to the case of a nonresonantly pumped condensate in the model 
%% with a separate reservoir field, two other groups have been investigating the post-quench dynamics described by the alternative
%% models of the stochastic Gross-Pitaevskii equation~\cite{Nick} and the resonantly pumped optical parametric 
%% oscillator system~\cite{Marzena}. In both cases, results indicate that the scaling law~(\ref{XYlaw}) holds also
%% in these systems.

We thank Alejandro Zamora, Marzena Szyma\'nska, and Nikolaos Proukakis for stimulating and valuable discussions. 
We acknowledge support from National Science Center, Poland grants DEC-2011/01/D/ST3/00482 and 2015/17/B/ST3/02273. 

\appendix %%%%%%%%%%%%%%%%%%%%%%%%%%%%%%%%%%%%%%%%%%%%%%%%%%%%%%%%%%% Appendices

\section{Phase ordering with hard-wall boundary conditions}

\begin{figure}[tbp]
 \centering
 \includegraphics[width=0.85\linewidth]{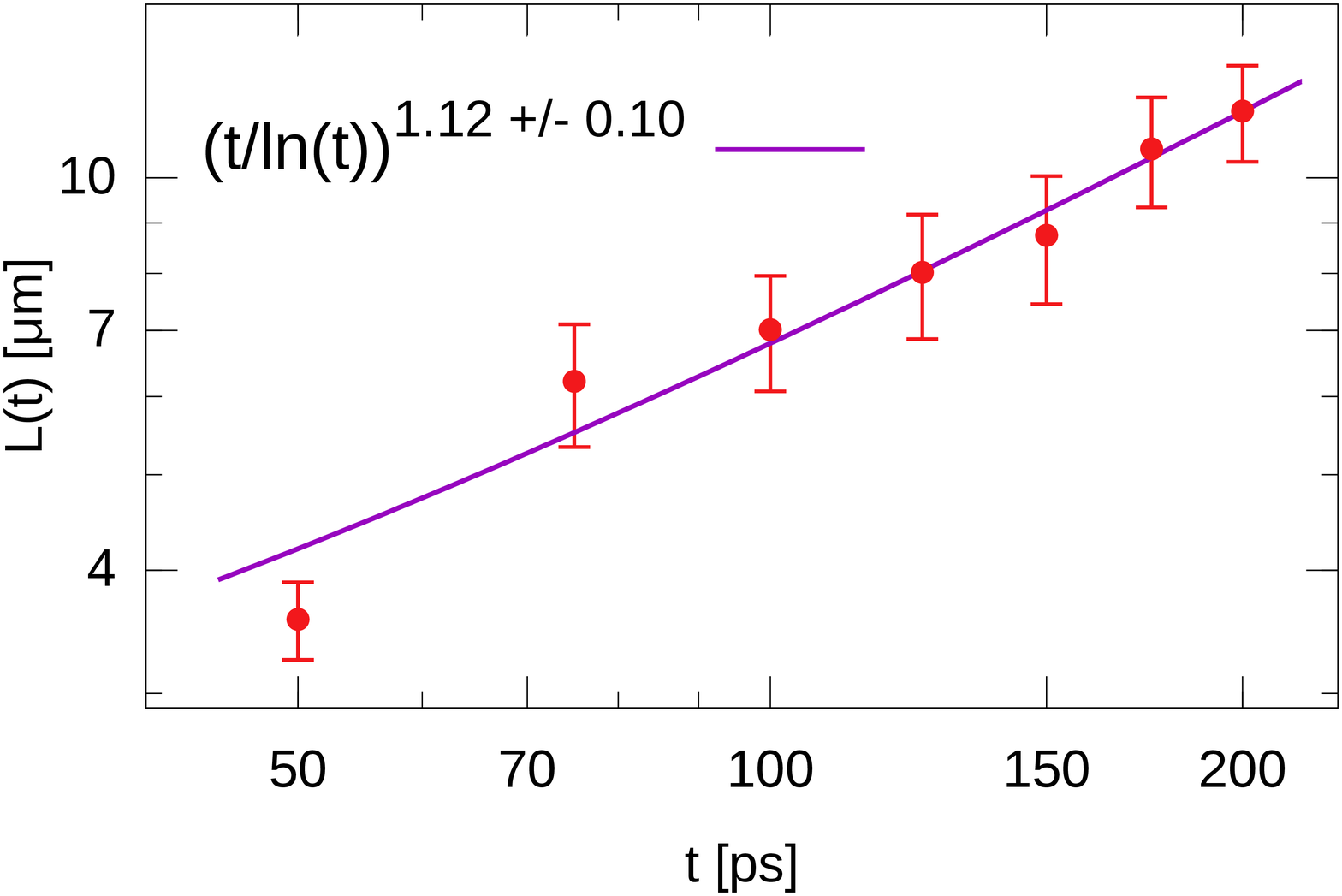}
 
 \caption{Time dependence of the length scale $L(t)$ in the case of a high quality sample with increased polariton lifetime, 
as in Fig.~\ref{fig:scalings}(a), with Dirichlet (hard-wall) boundary conditions.}
 \label{fig:Dirichlet}
 \end{figure}

In Fig.~\ref{fig:Dirichlet} we show the evolution of the length scale in the case with Dirichlet boundary conditions for the condensate wave function $\psi(\er,t)$. The obtained fit with dynamical exponent $1/z=1.12 \pm 0.10$ agrees with the one obtained in the case of periodic boundary conditions, see Fig.~4~(b) in the main text. Note that error bars are larger than in the periodic boundary case solely due to the way we perform averaging of the correlation function in this case.
As the condensate density always tends to zero close to the Dirichlet boundaries, we are no longer able to average over spatial coordinates r as in Eq. (1) in the main text. Instead we calculate correlation function from $x$ and $-x$ points on the sample
\begin{align}
g^{(1)}(x,t) = \frac{1}{N}\langle \psi^*(-x,0,t) \psi(x,0,t) \rangle
\end{align}
The absence of averaging over spatial coordinates leads to larger variations in the estimated correlation functions and the length scale L(t). 

\section{Decay of the number of vortices}

\begin{figure}[tbp]
 \centering
 \includegraphics[width=0.85\linewidth]{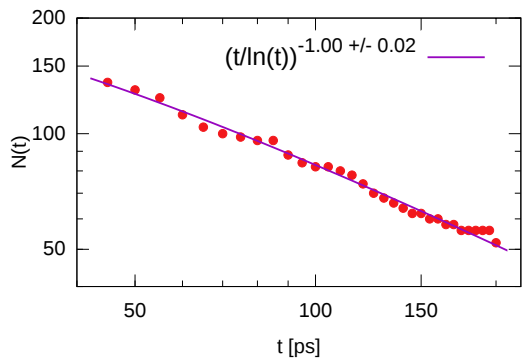}
 
 \caption{Time dependence of the vortex number $N(t)$ presented in bi-logarithmic scale, in the case of a low quality sample with short polariton lifetime (as in Fig.~4~(a) in the main text). The data corresponds to a single simulation.}
 \label{fig:VortexNumber}
\end{figure}

We used two automated methods of vortex counting, one based on the counting of local density dips, and the other based on calculation of local phase winding around a particular point on the grid. After the initial stage of evolution, when the condensate density is relatively high, the vortices are well defined and the two methods give the same results. The evolution of the vortex number is presented in Fig.~\ref{fig:VortexNumber}. It is in excellent agreement with the scaling of correlation function, taking into account that the number of vortices scales as $N\sim t^{-d/z}$ with $d=2$ being the number of dimensions and $z=2$.

\section{The effect of the logarithmic correction}

\begin{figure}
\includegraphics[width=6.5cm]{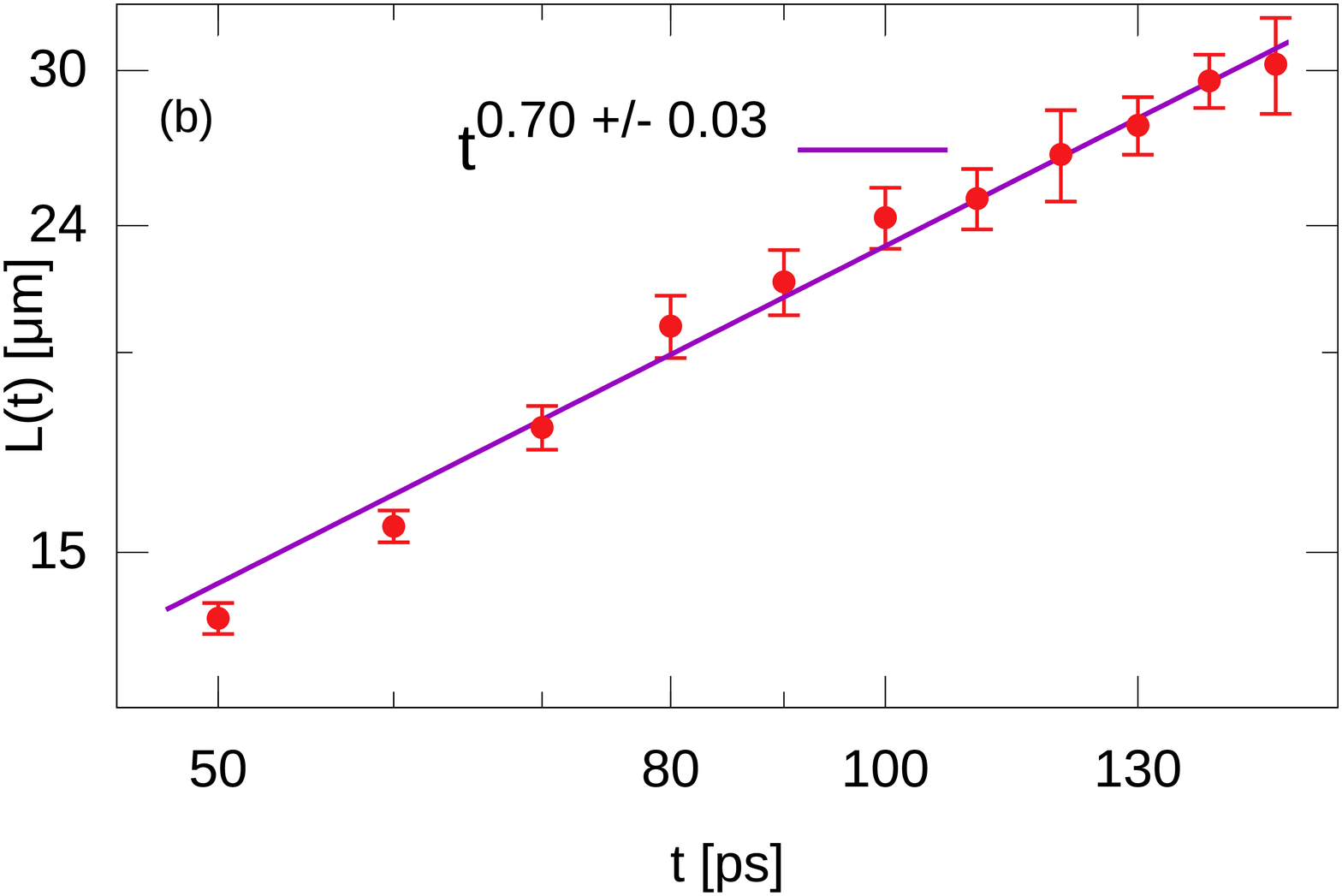}
\includegraphics[width=6.5cm]{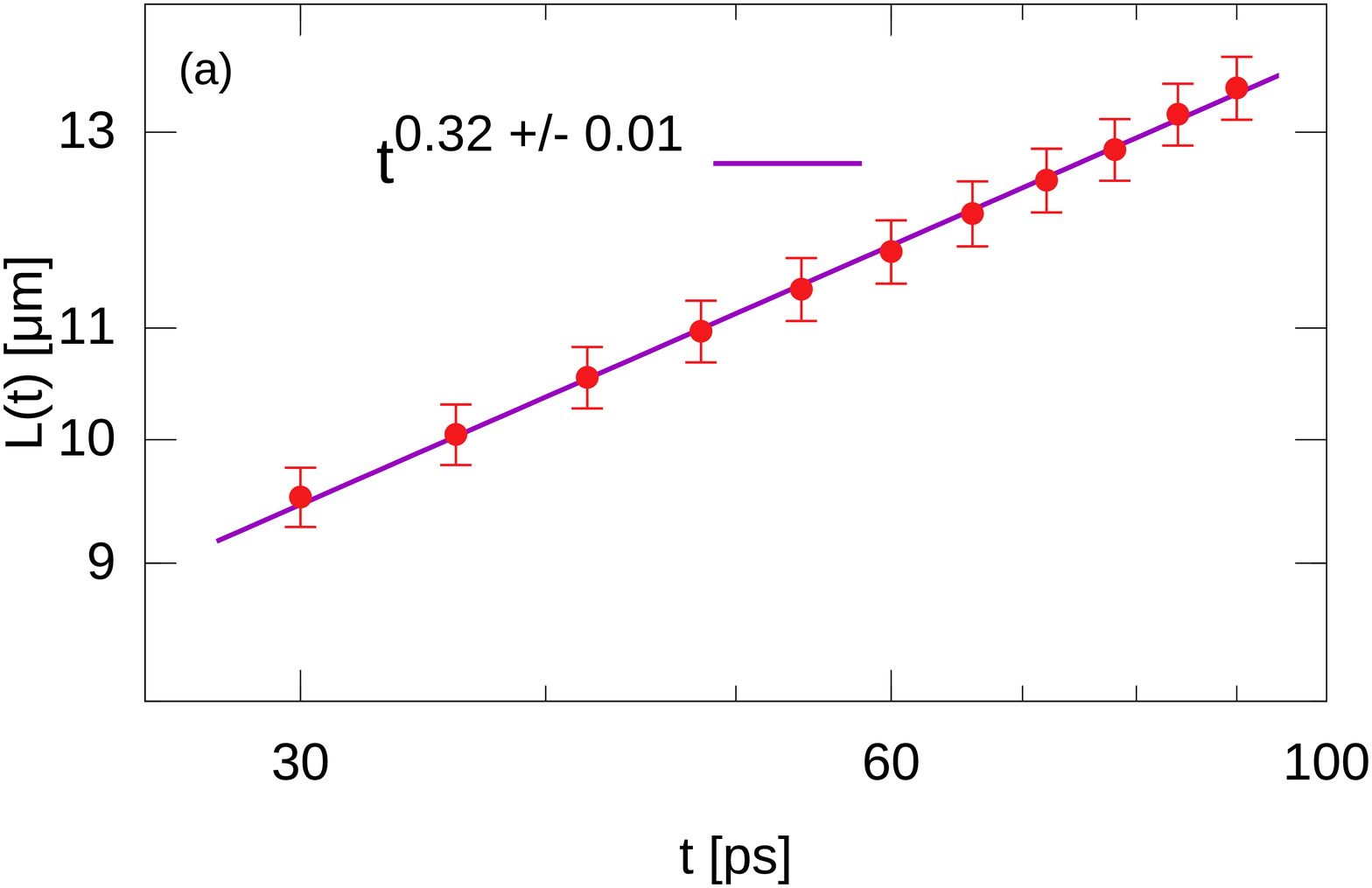}
\caption{Same as Fig.~\ref{fig:scalings}, but theoretical fits without the logarithmic correction.
}
\label{fig:scalings_noln}
\end{figure}

In Figure~\ref{fig:scalings_noln} we show the result of fitting the numerical data to the pure algebraic function without 
the logarithmic correction. The fit is also very good, but the value of the scaling exponents do not agree with theoretical 
predictions, Eq.~(\ref{XYlaw}). This is very similar to the situation described in the XY model~\cite{Bray_XYModel}, 
where this issue was discussed at length. 
The logarithmic correction effectively changes the slope of the fit, but does not result in significant bending 
in the bi-logarithmic scale. Such bending could be observable at early times t, but in this limit the universal scaling is not valid.


\begin{thebibliography}{99}

\bibitem{KibbleZurek} T. W. B. Kibble, J. Phys. A {\bf 9}, 1387 (1976);
W. H. \.Zurek, Nature (London) {\bf 317}, 505 (1985).

\bibitem{Bray_PhaseOrderingKinetics}
A. J. ~Bray
Adv. Phys. 43, 3 (1994)

\bibitem{Cahn} J. W. Cahn, Acta Metall. {\bf 9}, 795 (1961).

\bibitem{Furukawa_Scaling}
H. ~Furukawa,
Adv. Phys. 34, 6 (1985) 

\bibitem{Castellano_StatisticalSocial}
C. ~Castellano, S. ~Fortunato, and V. ~Loreto
Rev. Mod. Phys. 81, 591 (2009)


\bibitem{Sachdev_PhaseOrdering}
K. ~Damle, S. N. ~Majumdar, and S. ~Sachdev
Phys. Rev. A {\bf 54}, 5037 (1996)

\bibitem{Svistunov} Yu. M. Kagan, B.V. Svistunov, G.V. Shlyapnikov, Sov.~Phys.~JETP {\bf 75}, 387 (1992); 
Yu. Kagan, B.V. Svistunov, Sov.~Phys.~JETP {\bf 78}, 187 (1994); N. G. Berloff and B. V. Svistunov, Phys. Rev. A, {\bf 66}, 013603 (2002).

\bibitem{Moore_Spinor}
S. ~Mukerjee, C. ~Xu, and J. E. ~Moore
Phys. Rev. B. {\bf 76}, 104519 (2007)

\bibitem{Sarma_Coarsening}
J. ~Hofmann, S. S. ~Natu, and S. Das Sarma
Phys. Rev. Lett. {\bf 113}, 095702 (2014).

\bibitem{Kawaguchi_DomainGrowth}
K. ~Kudo, and Y. ~Kawaguchi
Phys. Rev. A {\bf 88}, 013630 (2013)

\bibitem{Kawaguchi_FerromagneticCoarsening}
K. Kudo and Y. Kawaguchi
Phys. Rev. A {\bf 91}, 053609 (2015)

\bibitem{Blakie_CoarseningSpin1}
L. A. Williamson, P. B. Blakie,
Phys. Rev. Lett.  {\bf 116}, 025301 (2016). 


\bibitem{Carusotto_QuantumFluids}
I. ~Carusotto,C. Ciuti
Rev. Mod. Phys. 85, 299 (2013)

\bibitem{Polaritons}
J.~J.~Hopfield, Phys. Rev. Lett. \textbf{112}, 1555 (1958);
C. ~Weisbuch, M. ~Nishioka, A. ~Ishikawa, and Y. ~Arakawa, Phys. Rev. Lett. \textbf{69}, 3314 (1992);
A. V. Kavokin, J. J. Baumberg, G. Malpuech, and F. P. Laussy,
{\it Microcavities} (Oxford University Press, Oxford, 2007).
%2
\bibitem{Kasprzak_BEC} J. Kasprzak, M. Richard, S. Kundermann, A. Baas, P. Jeambrun, J. M. J. Keeling, 
F. M. Marchetti, M. H. Szyma\'nska, R. Andr\'e, J. L. Staehli \textit{et al.}, 
Nature (London) {\bf 443}, 409 (2006).
%3

\bibitem{Deveaud_VortexDynamics} K. G. Lagoudakis, F. Manni, B. Pietka, M. Wouters, T. C. H. Liew, V. Savona, A. V. Kavokin,
R. Andr\'e, and B. Deveaud-Pl\'edran, Phys. Rev. Lett. {\bf 106}, 115301 (2011).

\bibitem{Matuszewski_UniversalityPolaritons}
M. Matuszewski and E. Witkowska,
Phys. Rev. B.  {\bf 89}, 155318 (2014).

\bibitem{Liew_InstabilityInduced}
T. C. H. ~Liew, O. A. ~Egorov, M. ~Matuszewski, O. ~Kyriienko, X. ~Ma, and E. A. ~Ostrovskaya
Phys. Rev. B. {\bf 91}, 085413 (2015)

\bibitem{Vina_VorticesCoherently}
D. Sanvitto,	 F. M. Marchetti, M. H. Szymañska, G. Tosi, M. Baudisch, F. P. Laussy, D. N. Krizhanovskii, M. S. Skolnick, L. Marrucci, A. Lemaitre, J. Bloch, C. Tejedor and L. Vina
Nature Physics {\bf 6},527–533 (2010)

\bibitem{Bramati_AngularVortexChain} T. Boulier, H.~Ter{\c c}as, D.~D.~Solnyshkov, Q.~Glorieux, E.~Giacobino, 
G.~Malpuech, and A.~Bramati,  Sci. Rep.  {\bf 5}, 9230 (2015). 

\bibitem{Deveaud_HydrodynamicVortices}
G. Nardin, G. Grosso, Y. Leger, B. Pietka, F. Morier-Genoud
Nature Physics 7, 635–641 (2011)

\bibitem{Deveaud_VortexStreets}
G. Grosso, G. Nardin, F. Morier-Genoud, Y. Leger, and B. Deveaud-Pledran
Phys. Rev. Lett. {\bf 107}, 245301 (2011)

\bibitem{Fraser_VortexAntivortex}
M. D. Fraser, G. Roumpos, and Y. Yamamoto
New J. Phys. {\bf 11}, 113048 (2009).

\bibitem{Rubo_WarpingVortices}
M. ~Toledo-Solano, M. E. ~Mora-Ramos, A. ~Figueroa, Y. G. ~Rubo
Phys. Rev. B.  {\bf 89}, 035308 (2014).

\bibitem{Skolnick_InteractionsOnVortices}
D. N. Krizhanovskii, D. M. Whittaker, R. A. Bradley, K. Guda, D. Sarkar, D. Sanvitto, L. Vina, E. Cerda, P. Santos,
K. Biermann, R. Hey, and M. S. Skolnick,
Phys. Rev. Lett. {\bf 104}, 126402 (2010)

\bibitem{Szymanska_TOPOVortices}
M. H. Szymanska, F. M. Marchetti, and D. Sanvitto
Phys. Rev. Lett. {\bf 105}, 236402 (2010)

\bibitem{Yamamoto_VortexPair}
G. Roumpos, M. D. Fraser, A. Loffler, S. Hofling, A. Forchel and Y. Yamamoto
Nature Physics {\bf 7}, 129–133 (2011)

\bibitem{Whittaker_TriggeredVortices}
F. M. Marchetti, M. H. Szymanska, C. Tejedor, and D. M. Whittaker
Phys. Rev. Lett. {\bf 105}, 063902 (2010)

\bibitem{Deveaud_SelectivePhotoexcitation} G.~Nardin, K.~G.~Lagoudakis, B.~Pietka, F.~Morier-Genoud, Y.~L{\'e}ger, 
and B.~Deveaud-Pl{\'e}dran,  Phys. Rev. B  {\bf 82}, 073303 (2010). 

\bibitem{Baumberg_GeometricallyLocked} Tosi, G., G.~Christmann, N.~G.~Berloff, P.~Tsotsis, T.~Gao, Z.~Hatzopoulos, 
P.~G.~Savvidis, and J.~J.~Baumberg,  Nature Communications  {\bf 3}, 1243 (2012). 

\bibitem{Vina_VortexAntivortex}
G. Tosi, F. M. Marchetti, D. Sanvitto, C. Anton, M. H. Szymanska, A. Berceanu, C. Tejedor, L. Marrucci, A. Lemaitre, J. Bloch, and L. Vina
Phys. Rev. Lett. {\bf 107}, 036401 (2011)

\bibitem{Sanvitto_AllOpticalControl} D.~Sanvitto, S.~Pigeon, A.~Amo, D.~Ballarini, M.~de Giorgi, I.~Carusotto, 
R.~Hivet, F.~Pisanello, V.~G.~Sala, P.~S.~S.~Guimaraes, R.~Houdr{\'e}, 
E.~Giacobino, C.~Ciuti, A.~Bramati, and G.~Gigli,  Nature Photon.  {\bf 
5}, 610 (2011); L.~Dominici, G.~Dagvadorj, J.~M.~Fellows, S.~Donati, D.~Ballarini, M.~De 
Giorgi, F.~M.~Marchetti, B.~Piccirillo, L.~Marrucci, A.~Bramati, G.~Gigli, 
M.~H.~Szyma{\'n}ska, and D.~Sanvitto,  
Science Advances {\bf 1}, e1500807 (2015). 

\bibitem{Bramati_MergingVortices}
E. Cancellieri, T. Boulier, R. Hivet, D. Ballarini, D. Sanvitto, M. H. Szymanska, C. Ciuti,
E. Giacobino, and A. Bramati,
Phys. Rev. B.  {\bf 90}, 214518 (2014).

\bibitem{Wouters_ExcitationSpectrum} M. Wouters and I. Carusotto,
Phys. Rev. Lett. {\bf 99}, 140402 (2007).

\bibitem{Bobrovska_Adiabatic} N.~Bobrovska and M.~Matuszewski,  Phys. Rev. B  {\bf 92}, 035311 
(2015). 

\bibitem{Szymanska_NonequilibriumBKT} G.~Dagvadorj, J.~M.~Fellows, S.~Matyja{\'s}kiewicz, F.~M.~Marchetti, 
I.~Carusotto, and M.~H.~Szyma{\'n}ska,  Phys. Rev. X  {\bf 5}, 041028 (2015). 


\bibitem{Pargellis_DefectDynamics}
A. N. Pargellis, P. Finn, J. W. Goodby, P. Panizza, B. Yurke and P. E. Cladis
Phys. Rev. A {\bf 46}, 7765 (1992).

\bibitem{Yamamoto_PNAS}
G.~Roumpos, M.~Lohse, W.~H.~Nitsche, J.~Keeling, M.~H.~Szyma{\'n}ska, 
P.~B.~Littlewood, A.~L{\"o}ffler, S.~H{\"o}fling, L.~Worschech, A.~Forchel, 
and Y.~Yamamoto,  Proc. Natl. Acad. Sci.  {\bf 109}, 6467 (2012). 

\bibitem{Yamamoto_AlgebraicBKT}
W.~H.~Nitsche, N.~Y.~Kim, G.~Roumpos, C.~Schneider, M.~Kamp, 
S.~H{\"o}fling, A.~Forchel, and Y.~Yamamoto,  Phys. Rev. B  {\bf 90}, 
205430 (2014). 

\bibitem{Sanvitto_TopologicalOrder}
D.~Caputo, D.~Ballarini, G.~Dagvadorj, C.~S{\'a}nchez Mu{\~n}oz, M.~De 
Giorgi, L.~Dominici, K.~West, L.~N.~Pfeiffer, G.~Gigli, F.~P.~Laussy, 
M.~H.~Szyma{\'n}ska, and D.~Sanvitto,  arXiv:1610.05737 (2016). 


\bibitem{Biroli_IsingAnnealing}
G. ~Biroli, L. F. ~Cugliandolo, and A. ~Sicilia
Phys. Rev. E. {\bf 81}, 050101(R) (2010)

\bibitem{Hohenberg_DynamicalCriticalPhenomena}
P. C. Hohenberg and B. I. Halperin
Rev. Mod. Phys. {\bf 49}, 435 (1977).

\bibitem{Furukawa_Droplet}
H. Furukawa
Phys. Rev. A {\bf 31}, 1103 (1985)

\bibitem{Yurke_CoarseningXY}
B. Yurke, A. N. Pargellis, T. Kovacs, and D. A. Huse
Phys. Rev. E {\bf 47}, 1525 (1993).

\bibitem{Bray_XYModel}
A. J. Bray, A. J. Briant, and D. K. Jervis
Phys. Rev. Lett. {\bf 84}, 1503 (2000)

\bibitem{BrayBlundell_OnModel}
R. E. Blundell and A. J. Bray
Phys. Rev. E. {\bf 49}, 4925 (1994)

\bibitem{Jelic_QuenchXYModel}
A. Jelic, L. F. Cugliandolo, J.~Stat.~Mech.~P02032 (2011). 

\bibitem{Bray_EnergyScalingApproach}
A. D. Rutenberg and A. J. Bray
Phys. Rev. E. {\bf 51}, 5499 (1995)

\bibitem{Bray_GrowthLaws}
A. J. Bray and A. D. Rutenberg
Phys. Rev. E. {\bf 49}, R27(R) (1994)

\bibitem{Wouters_ClassicalFields}
I. Carusotto and C. Ciuti, Phys. Rev. B {\bf 72}, 125335 (2005);
M. Wouters and V. Savona, Phys. Rev. B {\bf 79}, 165302 (2009).

\bibitem{Wouters_Superfluidity} Phys. Rev. Lett. {\bf 105}, 020602 (2010).

\bibitem{Wouters_EnergyRelaxation} M. Wouters, T. C. H. Liew, and V. Savona, Phys. Rev. B {\bf 82}, 245315 (2010).

\bibitem{Bobrovska_Stability}
N.~Bobrovska, E.~A. Ostrovskaya, and M.~Matuszewski,
Phys. Rev. B.  {\bf 90}, 205304 (2014).

\bibitem{Sieberer_DynamicalCritical}
L. M. Sieberer, S.~D. Huber, E. Altman, and S. Diehl,
Phys. Rev. Lett.  {\bf 110}, 195301 (2013).

\bibitem{Bloch_ExtendedCondensates}
E. Wertz, L. Ferrier, D. D. Solnyshkov, R. Johne, D. Sanvitto, A. Lema\^itre, I. Sagnes, R. Grousson, A. V. Kavokin, P. Senellart, \textit{et al.}, 
Nat. Phys. {\bf 6}, 860 (2010). % One Dimensional wire
%29
%31
\bibitem{Bloch_GapStates} D.~Tanese, H.~Flayac, D.~Solnyshkov, A.~Amo, A.~Lema{\^i}tre, 
E.~Galopin, R.~Braive, P.~Senellart, I.~Sagnes, G.~Malpuech, \textit{et al.},  Nat.
Commun.  {\bf 4}, 1749 (2013). 


\bibitem{Zurek_WhenSymmetryBreaks2D} A.~Yates and W.~H.~Zurek,  Phys. Rev. Lett.  {\bf 80}, 5477 (1998). 

\bibitem{Deutschlander_KZMColloids}
S. Deutschl\"ander, P. Dillmann, G. Maret, and P. Keim,
Proc. Natl. Acad. Sci,  {\bf 112},  6925 (2015).

\bibitem{Szymanska_NonequilibriumCondensation} M.~H. Szyma{\'n}ska, J.~Keeling, and P.~B.~Littlewood,  
Phys. Rev. Lett.  {\bf 96}, 230602 (2006). 

\bibitem{Diehl_TwoDimensionalSuperfluidity} E. Altman, L.~M.~Sieberer, L.~Chen, S.~Diehl, and J.~Toner,  
Phys Rev. X {\bf 5}, 011017 (2015). 


\bibitem{Fetter_vortices} A.~L.~Fetter, and A.~A.~Svidzinsky,  J. Phys. Condens. Matter
{\bf 13}, R135 (2001). 

\bibitem{Diehl_Duality} G. Wachtel, L. M. Sieberer, S. Diehl, and E. Altman, Phys. Rev. B {\bf 94}, 104520 (2016).

%% ---

%% \bibitem{KibbleZurek} T. W. B. Kibble, J. Phys. A {\bf 9}, 1387 (1976);
%% W. H. \.Zurek, Nature (London) {\bf 317}, 505 (1985).

%% \bibitem{Matuszewski_UniversalityPolaritons}
%% M. Matuszewski and E. Witkowska,
%% Phys. Rev. B.  {\bf 89}, 155318 (2014).

%% \bibitem{Liew_InstabilityInduced}
%% T. C. H. ~Liew, O. A. ~Egorov, M. ~Matuszewski, O. ~Kyriienko, X. ~Ma, and E. A. ~Ostrovskaya
%% Phys. Rev. B. {\bf 91}, 085413 (2015)

%% \bibitem{Bray_PhaseOrderingKinetics}
%% A. J. ~Bray
%% Adv. Phys. 43, 3 (1994)

%% \bibitem{Cahn} J. W. Cahn, Acta Metall. {\bf 9}, 795 (1961).

%% \bibitem{Furukawa_Scaling}
%% H. ~Furukawa
%% Adv. Phys. 34, 6 (1985) 

%% \bibitem{Castellano_StatisticalSocial}
%% C. ~Castellano, S. ~Fortunato, and V. ~Loreto
%% Rev. Mod. Phys. 81, 591 (2009)

%% \bibitem{Moore_Spinor}
%% S. ~Mukerjee, C. ~Xu, and J. E. ~Moore
%% Phys. Rev. B. {\bf 76}, 104519 (2007)

%% \bibitem{Sarma_Coarsening}
%% J. ~Hofmann, S. S. ~Natu, and S. Das Sarma
%% Phys. Rev. Lett. {\bf 113}, 095702 (2014).

%% \bibitem{Sachdev_PhaseOrdering}
%% K. ~Damle, S. N. ~Majumdar, and S. ~Sachdev
%% Phys. Rev. A {\bf 54}, 5037 (1996)

%% \bibitem{Furukawa_Droplet}
%% H. Furukawa
%% Phys. Rev. A {\bf 31}, 1103 (1985)

%% \bibitem{Kawaguchi_DomainGrowth}
%% K. ~Kudo, and Y. ~Kawaguchi
%% Phys. Rev. A {\bf 88}, 013630 (2013)

%% \bibitem{Kawaguchi_FerromagneticCoarsening}
%% K. Kudo and Y. Kawaguchi
%% Phys. Rev. A {\bf 91}, 053609 (2015)

%% \bibitem{Blakie_CoarseningSpin1}
%% L. A. Williamson, P. B. Blakie
%% arXiv:1504.06404(2015)

%% \bibitem{Biroli_IsingAnnealing}
%% G. ~Biroli, L. F. ~Cugliandolo, and A. ~Sicilia
%% Phys. Rev. E. {\bf 81}, 050101(R) (2010)

%% \bibitem{Carusotto_QuantumFluids}
%% I. ~Carusotto,C. Ciuti
%% Rev. Mod. Phys. 85, 299 (2013)


%% \bibitem{Polaritons}
%% J.~J.~Hopfield, Phys. Rev. Lett. \textbf{112}, 1555 (1958);
%% C. ~Weisbuch, M. ~Nishioka, A. ~Ishikawa, and Y. ~Arakawa, Phys. Rev. Lett. \textbf{69}, 3314 (1992);
%% A. V. Kavokin, J. J. Baumberg, G. Malpuech, and F. P. Laussy,
%% {\it Microcavities} (Oxford University Press, Oxford, 2007).
%% %2
%% \bibitem{Kasprzak_BEC} J. Kasprzak, M. Richard, S. Kundermann, A. Baas, P. Jeambrun, J. M. J. Keeling, 
%% F. M. Marchetti, M. H. Szyma\'nska, R. Andr\'e, J. L. Staehli \textit{et al.}, 
%% Nature (London) {\bf 443}, 409 (2006).
%% %3

%% \bibitem{Wouters_ExcitationSpectrum} M. Wouters and I. Carusotto,
%% Phys. Rev. Lett. {\bf 99}, 140402 (2007).

%% \bibitem{Deveaud_VortexDynamics} K. G. Lagoudakis, F. Manni, B. Pietka, M. Wouters, T. C. H. Liew, V. Savona, A. V. Kavokin,
%% R. Andr\'e, and B. Deveaud-Pl\'edran, Phys. Rev. Lett. {\bf 106}, 115301 (2011).

%% \bibitem{Wouters_ClassicalFields}
%% I. Carusotto and C. Ciuti, Phys. Rev. B {\bf 72}, 125335 (2005);
%% M. Wouters and V. Savona, Phys. Rev. B {\bf 79}, 165302 (2009).

%% \bibitem{Vina_VorticesCoherently}
%% D. Sanvitto,	 F. M. Marchetti, M. H. Szymañska, G. Tosi, M. Baudisch, F. P. Laussy, D. N. Krizhanovskii, M. S. Skolnick, L. Marrucci, A. Lemaitre, J. Bloch, C. Tejedor and L. Vina
%% Nature Physics {\bf 6},527–533 (2010)

%% \bibitem{Bramati_AngularVortexChain} T. Boulier, H.~Ter{\c c}as, D.~D.~Solnyshkov, Q.~Glorieux, E.~Giacobino, 
%% G.~Malpuech, and A.~Bramati,  Sci. Rep.  {\bf 5}, 9230 (2015). 

%% \bibitem{Bramati_MergingVortices}
%% E. Cancellieri, T. Boulier, R. Hivet, D. Ballarini, D. Sanvitto, M. H. Szymanska, C. Ciuti
%% Phys. Rev. B.  {\bf 90}, 214518 (2014).

%% \bibitem{Deveaud_HydrodynamicVortices}
%% G. Nardin, G. Grosso, Y. Leger, B. Pietka, F. Morier-Genoud
%% Nature Physics 7, 635–641 (2011)

%% \bibitem{Deveaud_VortexStreets}
%% G. Grosso, G. Nardin, F. Morier-Genoud, Y. Leger, and B. Deveaud-Pledran
%% Phys. Rev. Lett. {\bf 107}, 245301 (2011)

%% \bibitem{Fraser_VortexAntivortex}
%% M. D. Fraser, G. Roumpos, and Y. Yamamoto
%% New J. Phys. {\bf 11}, 113048 (2009).

%% \bibitem{Rubo_WarpingVortices}
%% M. ~Toledo-Solano, M. E. ~Mora-Ramos, A. ~Figueroa, Y. G. ~Rubo
%% Phys. Rev. B.  {\bf 89}, 035308 (2014).

%% \bibitem{Skolnick_InteractionsOnVortices}
%% D. N. Krizhanovskii, D. M. Whittaker, R. A. Bradley, K. Guda, D. Sarkar, D. Sanvitto, L. Vina, E. Cerda, P. Santos
%% Phys. Rev. Lett. {\bf 104}, 126402 (2010)

%% \bibitem{Szymanska_TOPOVortices}
%% M. H. Szymanska, F. M. Marchetti, and D. Sanvitto
%% Phys. Rev. Lett. {\bf 105}, 236402 (2010)

%% \bibitem{Yamamoto_VortexPair}
%% G. Roumpos,	M. D. Fraser, A. Loffler, S. Hofling, A. Forchel	and Y. Yamamoto
%% Nature Physics {\bf 7}, 129–133 (2011)

%% \bibitem{Whittaker_TriggeredVortices}
%% F. M. Marchetti, M. H. Szymanska, C. Tejedor, and D. M. Whittaker
%% Phys. Rev. Lett. {\bf 105}, 063902 (2010)


%% \bibitem{Vina_VortexAntivortex}
%% G. Tosi, F. M. Marchetti, D. Sanvitto, C. Anton, M. H. Szymanska, A. Berceanu, C. Tejedor, L. Marrucci, A. Lemaitre, J. Bloch, and L. Vina
%% Phys. Rev. Lett. {\bf 107}, 036401 (2011)

%% \bibitem{Cahn} J. W. Cahn, Acta Metall. {\bf 9}, 795 (1961).

%% \bibitem{Sanvitto_AllOpticalControl} D.~Sanvitto, S.~Pigeon, A.~Amo, D.~Ballarini, M.~de Giorgi, I.~Carusotto, 
%% R.~Hivet, F.~Pisanello, V.~G.~Sala, P.~S.~S.~Guimaraes, R.~Houdr{\'e}, 
%% E.~Giacobino, C.~Ciuti, A.~Bramati, and G.~Gigli,  Nature Photon.  {\bf 
%% 5}, 610 (2011); L.~Dominici, G.~Dagvadorj, J.~M.~Fellows, S.~Donati, D.~Ballarini, M.~De 
%% Giorgi, F.~M.~Marchetti, B.~Piccirillo, L.~Marrucci, A.~Bramati, G.~Gigli, 
%% M.~H.~Szyma{\'n}ska, and D.~Sanvitto,  
%% arXiv:1403.0487 (2014). 

%% \bibitem{Bramati_AngularVortexChain} T. Boulier, H.~Ter{\c c}as, D.~D.~Solnyshkov, Q.~Glorieux, E.~Giacobino, 
%% G.~Malpuech, and A.~Bramati,  Sci. Rep.  {\bf 5}, 9230 (2015). 

%% \bibitem{Wouters_ExcitationSpectrum} M. Wouters and I. Carusotto,
%% Phys. Rev. Lett. {\bf 99}, 140402 (2007).

%% \bibitem{Pargellis_DefectDynamics}
%% A. N. Pargellis, P. Finn, J. W. Goodby, P. Panizza, B. Yurke and P. E. Cladis
%% Phys. Rev. A {\bf 46}, 7765 (1992).


%% \bibitem{Hohenberg_DynamicalCriticalPhenomena}
%% P. C. Hohenberg and B. I. Halperin
%% Rev. Mod. Phys. {\bf 49}, 435 (1977).

%% \bibitem{Yurke_CoarseningXY}
%% B. Yurke, A. N. Pargellis, T. Kovacs, and D. A. Huse
%% Phys. Rev. E {\bf 47}, 1525 (1993).

%% \bibitem{Bray_XYModel}
%% A. J. Bray, A. J. Briant, and D. K. Jervis
%% Phys. Rev. Lett. {\bf 84}, 1503 (2000)

%% \bibitem{BrayBlundell_OnModel}
%% R. E. Blundell and A. J. Bray
%% Phys. Rev. E. {\bf 49}, 4925 (1994)

%% \bibitem{Jelic_QuenchXYModel}
%% A. Jelic, L. F. Cugliandolo
%% arXiv:1012.0417(2010)

%% \bibitem{Bray_EnergyScalingApproach}
%% A. D. Rutenberg and A. J. Bray
%% Phys. Rev. E. {\bf 51}, 5499 (1995)

%% \bibitem{Bray_GrowthLaws}
%% A. J. Bray and A. D. Rutenberg
%% Phys. Rev. E. {\bf 49}, R27(R) (1994)

%% \bibitem{KibbleZurek} T. W. B. Kibble, J. Phys. A {\bf 9}, 1387 (1976);
%% W. H. \.Zurek, Nature (London) {\bf 317}, 505 (1985).

%% \bibitem{Wouters_ClassicalFields}
%% I. Carusotto and C. Ciuti, Phys. Rev. B {\bf 72}, 125335 (2005);
%% M. Wouters and V. Savona, Phys. Rev. B {\bf 79}, 165302 (2009).

%% \bibitem{Matuszewski_UniversalityPolaritons}
%% M. Matuszewski and E. Witkowska,
%% Phys. Rev. B.  {\bf 89}, 155318 (2014).

%% \bibitem{Bloch_ExtendedCondensates}
%% E. Wertz, L. Ferrier, D. D. Solnyshkov, R. Johne, D. Sanvitto, A. Lema\^itre, I. Sagnes, R. Grousson, A. V. Kavokin, P. Senellart, \textit{et al.}, 
%% Nat. Phys. {\bf 6}, 860 (2010). % One Dimensional wire
%% %29
%% \bibitem{Bobrovska_Stability}
%% N.~Bobrovska, E.~A. Ostrovskaya, and M.~Matuszewski,
%% Phys. Rev. B.  {\bf 90}, 205304 (2014).

%% \bibitem{Sieberer_DynamicalCritical}
%% L. M. Sieberer, S.~D. Huber, E. Altman, and S. Diehl,
%% Phys. Rev. Lett.  {\bf 110}, 195301 (2013).

%% \bibitem{Bloch_ExtendedCondensates}
%% E. Wertz, L. Ferrier, D. D. Solnyshkov, R. Johne, D. Sanvitto, A. Lema\^itre, I. Sagnes, R. Grousson, A. V. Kavokin, P. Senellart, \textit{et al.}, 
%% Nat. Phys. {\bf 6}, 860 (2010). % One Dimensional wire
%% %29
%% %31
%% \bibitem{Bloch_GapStates} D.~Tanese, H.~Flayac, D.~Solnyshkov, A.~Amo, A.~Lema{\^i}tre, 
%% E.~Galopin, R.~Braive, P.~Senellart, I.~Sagnes, G.~Malpuech, \textit{et al.},  Nat.
%% Commun.  {\bf 4}, 1749 (2013). 



\end{thebibliography}
\end{document}